\documentclass{elsart}

\usepackage{natbib}

\usepackage{graphicx}

\usepackage{amssymb}
\usepackage{amsmath}

\begin{document}

\begin{frontmatter}

\title{The Gas Pixel Detector as a Solar X-Ray Polarimeter and Imager} 

\author{Sergio Fabiani\corauthref{cor}}
\address{INAF-IASF Roma, Via del Fosso del Cavaliere 100, 00133 Rome, Italy \\ Universit\`a di Roma 'Tor Vergata', Dipartimento di Fisica,
      Via della Ricerca Scientifica 1, 00133 Rome, Italy}
\corauth[cor]{Corresponding author}
\ead{sergio.fabiani@iasf-roma.inaf.it}

\author{Enrico Costa}
\address{INAF-IASF Roma, Via del Fosso del Cavaliere 100, 00133 Rome, Italy}

\author{Ronaldo Bellazzini}
\address{Istituto Nazionale di Fisica Nucleare, Largo B. Pontecorvo 3, I-56127 Pisa, Italy}

\author{Alessandro Brez}
\address{Istituto Nazionale di Fisica Nucleare, Largo B. Pontecorvo 3, I-56127 Pisa, Italy}

\author{Sergio Di Cosimo}
\address{INAF-IASF Roma, Via del Fosso del Cavaliere 100, 00133 Rome, Italy}

\author{Francesco Lazzarotto}
\address{INAF-IASF Roma, Via del Fosso del Cavaliere 100, 00133 Rome, Italy}

\author{Fabio Muleri}
\address{INAF-IASF Roma, Via del Fosso del Cavaliere 100, 00133 Rome, Italy}

\author{Alda Rubini}
\address{INAF-IASF Roma, Via del Fosso del Cavaliere 100, 00133 Rome, Italy}

\author{Paolo Soffitta}
\address{INAF-IASF Roma, Via del Fosso del Cavaliere 100, 00133 Rome, Italy}

\author{Gloria Spandre}
\address{Istituto Nazionale di Fisica Nucleare, Largo B. Pontecorvo 3, I-56127 Pisa, Italy}

\begin{abstract}

The Sun is the nearest astrophysical source with a very intense emission in the X-ray band. 
The study of energetic events, such as solar flares, can help us to understand the behaviour of the magnetic field of our star. 
There are in the literature numerous studies published about polarization predictions, for a wide range of solar flares models 
involving the emission from thermal and/or non-thermal processes, but observations in the X-ray band have never been exhaustive.

The Gas Pixel Detector (GPD) was designed to achieve X-ray polarimetric measurements as well as X-ray images for far astrophysical sources.
Here we present the possibility to employ this instrument for the observation of our Sun in the X-ray band.

\end{abstract}


\end{frontmatter}

\parindent=0.5 cm

\section{Introduction}

The Gas Pixel Detector (GPD) has been developed by the INFN and the IASF-Roma / INAF Italian research institutes. 
The GPD has been designed to achieve X-ray polarimetric measurements as well as X-ray images for far astrophysical sources. 
It has a good spectroscopic sensitivity thanks to an energy resolution of $\sim$ 20$\%$ at 6 keV and it allows also to 
perform accurate timing measurements.

Differently from all the other kinds of X-ray polarimeters, it doesn't need rotation.
The GPD was born to be employed at the focal plain of X-ray telescopes. Originally  the working energy  band was 2$\div$10 keV. 
Recently the possibility to extend the employment to higher energies has been taken into account \citep{Mul06, Sof10}. 
Here we want to discuss the use of this instrument, with an array configuration in association with 
a coded mask aperture, or at the focal plane of an X-ray telescope, to observe X-rays from solar flares.

There are in the literature measurements of solar flares X-ray polarization, but they are characterized by high uncertainties and sometimes very low significance \citep{Tin70, Tin72, Tin76, Zit06, Bog06, Su06}. For this, polarimetry of X-ray solar flare is still an open field which need to be deeply investigated.

\section{The Gas Pixel Detector (GPD)}

\begin{figure}
\begin{center}
\includegraphics*[width = 10 cm , angle = 0]{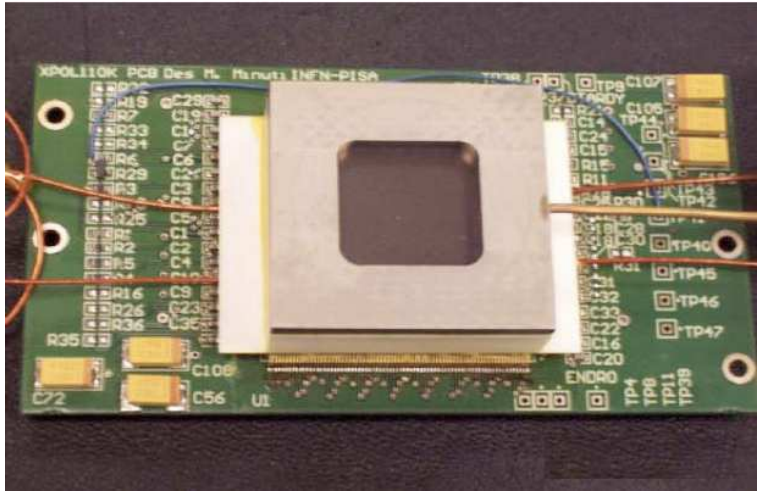}
\end{center}
\caption{View of the GPD.}\label{fig:figure1}
\end{figure}

\begin{figure}
\begin{center}
\includegraphics*[width = 10 cm , angle = 0]{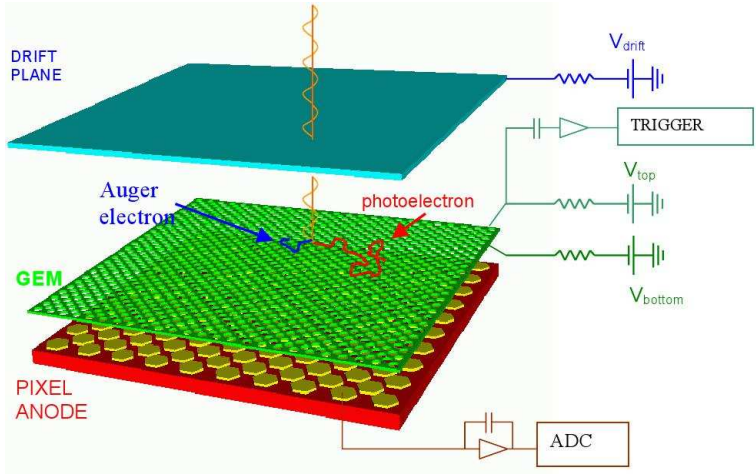}
\end{center}
\caption{Scheme of the Gas Pixel Detector.}\label{fig:figure2}
\end{figure}

The GPD (see Fig.\ref{fig:figure1}) is a gas detector which measures the polarization of X-ray radiation by exploiting the dependence of photoelectric cross section with respect to the direction of polarization of the detected photons. \citep{Bel07}.
The scheme of the GPD is shown in Fig.\ref{fig:figure2}. Photons enter into a cell filled with a gas mixture, passing through a thin beryllium window. They are absorbed via photoelectric effect by the gas.

\begin{figure}
\begin{center}
\includegraphics*[width= 10 cm , angle = 0]{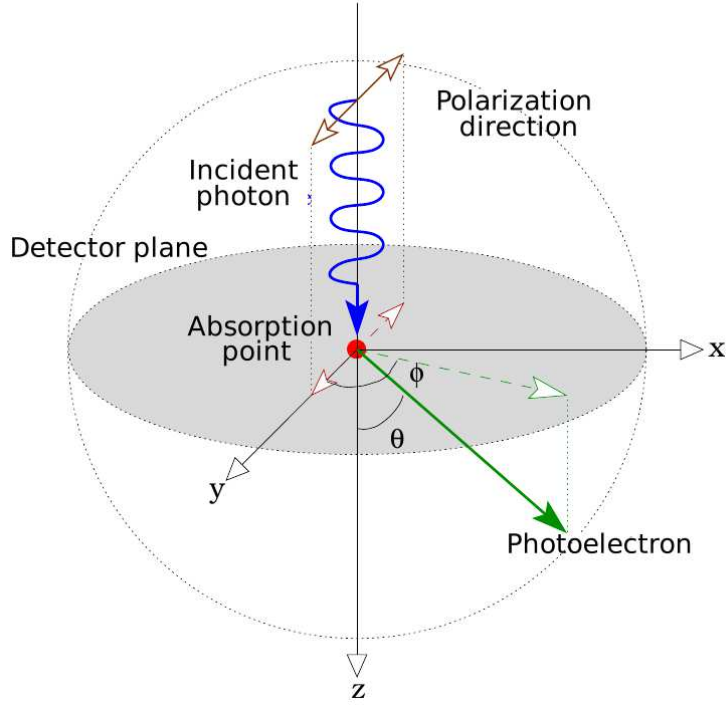}
\end{center}
\caption{System of reference to describe the photoelectric interaction.}\label{fig:figure3}
\end{figure}

To describe briefly the geometry of the interactions within the detector it is useful to introduce a Cartesian system of reference $xyz$ (see Fig.\ref{fig:figure3}). The origin is in the photon absorption point and the $y$ axis is aligned with the direction of polarization of the absorbed photon itself. Let be $\theta$, the angle between the arrival direction of the photon and the direction of ejection of the photoelectron, and $\phi$, the angle between the polarization direction and the direction of ejection of the photoelectron projected on the plane $xy$. 

The angular dependence of the photoelectric differential cross section is expressed in the Eq.\ref{eq:1}, which tells us that the photoelectron is emitted with high probability in the direction of the electric field of the absorbed photon ($\phi=0$). 
  \begin{equation}
    \label{eq:1}
 {%
   \frac{d\sigma_{ph}}{d\Omega} \propto sin^2{\theta} \cos^2{\phi} 
    }
  \overfullrule 5pt
  \mathindent\linewidth\relax
  \advance\mathindent-259pt
  \end{equation}
Such a photoelectron propagates and loses its energy ionizing gas atoms. It also suffers scattering events hitting charges in the nuclei. 
Thus electron-ion pairs are produced along the photoelectron path.  
The Bethe-Block formula describes the energy lost for ionization interactions. For electron energy $E_e$ of the order of the keV the energy lost follows the relation:
  \begin{equation}
    \label{eq:2}
 {%
-\frac{dE_e}{ds}=\frac{Z}{E_e}
    }
  \overfullrule 5pt
  \mathindent\linewidth\relax
  \advance\mathindent-259pt
  \end{equation}
where $Z$ is the atomic number of the gas, and $s$ is the path travelled. Eq.\ref{eq:2} tells us that a larger amount of energy is lost at the end of the path when the photoelectron is stopping. This lose of energy produces the major number of ionization electrons (the Bragg peak).

These ionization electrons are drifted and amplified by the Gas Electron Multiplier (GEM) and eventually collected on a fine sub-divided pixel detector located to the opposite site of the entrance window. 
The GEM is an electrode consisting of two metal layers separated by a thin insulator, etched with a regular hexagonal pattern of holes. 
The difference of potential established between the upper and the lower face of the GEM allows for charge multiplication.

At the present the core of the detector is 1.5 $\times$ 1.5 cm$^2$ and weights only 50 g.  The chip integrates more than 16.5 million  transistors. 
It has an active area of 105’600 pixels of 50 $\mu$m organized in a honeycomb matrix 300 $\times$ 352.  It is a self triggered system able to select by itself the pixel region which collects the charge produced by a track to perform charge download from such a small portion of the chip. 
The whole detector has a dimension of 14 $\times$ 19 $\times$ 7 cm$^3$ and weights only 1.6 kg, including the processing and control electronics.

By changing the pressure, the gas mixture and the depth of the cell, it is possible to select properly the energy band within which the instrument can work in the range of about 2$\div$35 keV. This is the wide band in which the photoelectric absorption is effective in gas.

\section{Tracks Analysis}

\begin{figure}
\begin{center}
\includegraphics*[width = 10 cm , angle = 0]{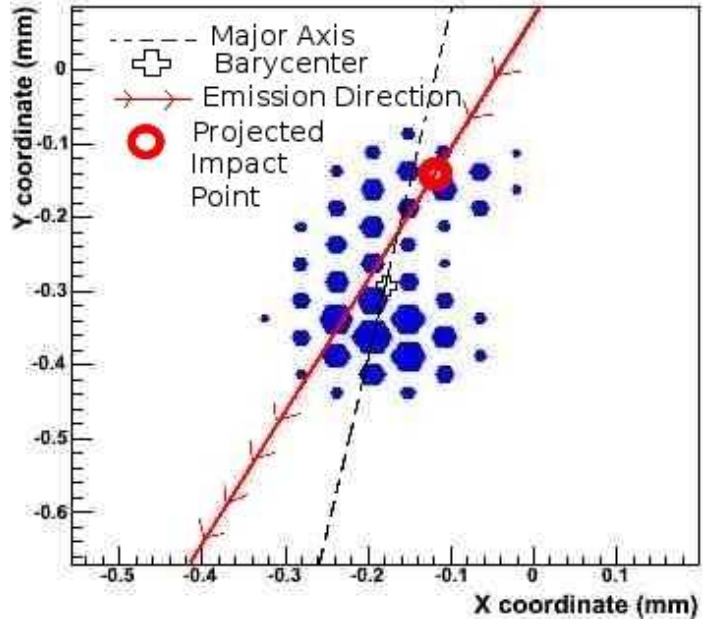}
\end{center}
\caption{Charge distribution given by the projection of a photoelectron track on the pixel plane from a 4.5 keV photon. The red solid line is the photoelectron emission direction and the thick circle is the projected absorption point. The gas mixture is 100$\%$ DME at 0.8 bar of pressure.}\label{fig:figure4}
\end{figure}

The photoelectron track is identified by the projected charge distribution on the pixel plane (see Fig.\ref{fig:figure4}). 
The analysis of the statistical momenta allows to evaluate some properties of this distribution such as: the barycentre, the major axis and most importantly the projection on the pixel plane of the point of absorption of the X-ray photon and the direction of ejection of the photoelectron. The key point is the distribution of the ejection direction and its relation with the polarization direction of the photon.
 
\begin{figure}
\begin{center}
\includegraphics*[width = 10 cm , angle = 0]{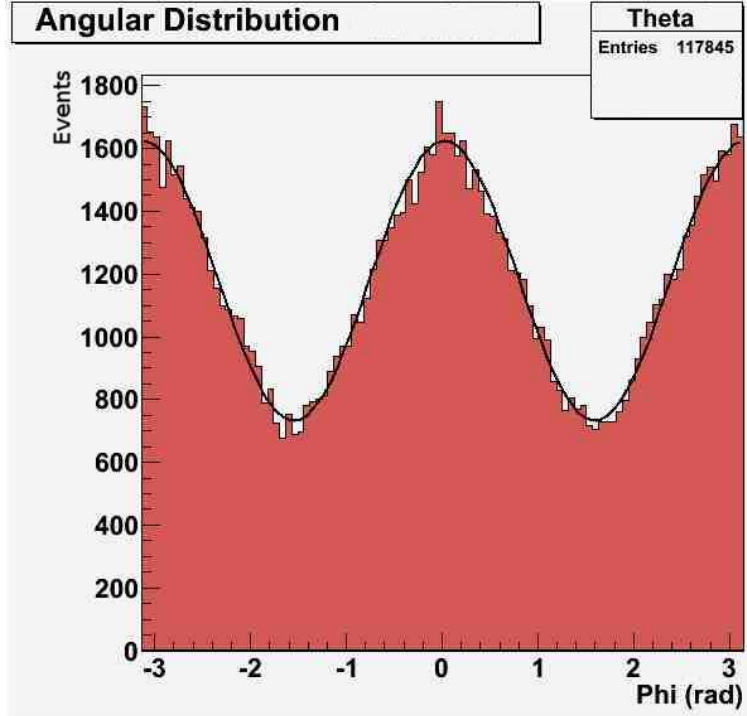}
\end{center}
\caption{Modulation histogram of the emission directions obtained analysing a beam of 100$\%$ polarized radiation at 4.5 keV absorbed within a gas cell 1 cm thick filled with 100$\%$ DME gas mixture at 0.8 bar of pressure. The modulation factor is (37.78$\pm$0.29)$\%$. No cuts on tracks shape are applied.}\label{fig:figure5}
\end{figure}

Thus a beam of polarized radiation produces a $\cos^2 \phi$ modulation pattern (the so called  modulation curve) from the distribution of the $\phi$ angles of the projections of the ejection directions on the pixel plane (see Fig.\ref{fig:figure5}). This because the direction of polarization corresponds to a preferential direction of ejection for the photoelectrons. 

From the amplitude and the phase angle of the modulation curve it is possible to obtain respectively the degree and the angle of polarization of the absorbed radiation.

Since by the analysis is possible to infer the projection on the pixel plane of the photon absorption point located in the gas cell, the GPD is a position sensitive instrument. 
Moreover the $\sim$150 $\mu$m of spatial resolution (given by tracks analysis) allows a good imaging capability.

\section{Solar Flare X-Ray Emission}

Solar flares affect all layers of the solar atmosphere in active regions.
They are thought to be generated by  magnetic reconnection consisting in a rearrangement of oppositely directed magnetic fields brought together. There are clear evidences that the reconnection produces a sudden release of energy stored in the original fields configuration and a violent particles acceleration takes place. Particles precipitate downwards to the lower solar atmosphere, while depositing their energy in the ambient plasma that heats \citep{Brown71, Hudson72, Lin76}. Non-thermal Hard X-Rays (HXRs) emission results from electrons that slow down producing bremsstrahlung radiation. 
However there are events for which Soft X-Rays (SHRs) emission produced by thermal heating from the reconnection site is not negligible,
contradicting the scheme according to which the atmospheric heating is a consequence of particles slowdown \citep{Per87, Flud95, Batt09}.

Solar flares emission is characterized by thermal emission from hot plasma (at low energies $<$10 keV) and non-thermal 
bremsstrahlung from electrons (at high energies $>$20 keV), resulting in a complex  spectrum which is moreover strongly characterized by lines emission up to about 7 keV \citep{Dosh02}.      
SXRs are the characteristic emission from the flare loop, whereas the HXRs are the typical emission from the loop foot-points, but also from the region interested by the reconnection of the magnetic field at the loop top. 
Such emissions evolve with time. Typically a flare is announced by the increase of the SXR emission, followed by the HXR activity. Moreover the spectral shape changes with time, passing through hardening and softening phases.  

Polarimetry should be performed with adequate time of accumulation per event and with sufficient angular resolution (at least 20'' to separate footpoints and loop-top sources). This to prevent the cancellation of the polarization vector due to possible time variations and to spatial symmetry. The capability to integrate events not only for the whole live time of the flare, but also during a single HXR peak should be a good goal for polarimetry measurements.

Polarization depends both on electrons beaming properties and on viewing angles. 
Polarized signals are expected from many models of non-thermal emission, associated to anisotropic distributions of electrons that are accelerated in ordered magnetic fields \citep{Brw74, LaP77, EmsVla80, LeacPet83, Zha95, Cha96}. 
Also thermal radiation is expected to be polarized, but only  at the level of some per cent, due to possible anisotropies in the electron distribution function \citep{EmsBro80}.
Moreover the backscattering of radiation on lower levels of solar atmosphere can induce polarization and modify 
polarization properties of reflected radiation \citep{BaiRa78}.

A recent work by \citep{Zha10} point out that the HXR bremsstrahlung emission from energetic beamed electrons can be significantly polarized. In their model the authors consider the injection of beamed electrons into a flaring atmosphere. 
They show a complex model taking into account magnetic mirroring, dues to magnetic field convergence, and return current, dues to the fact that the electric field carried by beam electrons causes preferential scattering around the backward direction to produce returning electrons with field directed in the opposite direction of the original one. These two phenomena may contribute heavily to the formation of an upward motion of particles.
The highest degree of polarization is expected for radiation emitted orthogonally with respect to the beam direction (aligned to the magnetic field). Thus flares located near the solar limb should appear more polarized, reaching degree of polarization of about 40$\%$ at 20 keV.
The degree of polarization of this non-thermal radiation is expected to decrease moving from low to high energies.

\section{Looking at the Sun with the GPD}

Polarimeters performances can be evaluated with the Minimum Detectable Polarization (MDP), that is the minimum polarization degree detectable at a significant confidence level. The MDP, for a 99$\%$ of confidence level is \citep{Weiss2010}:
  \begin{equation}
    \label{eq:3}
 {%
MDP(99\%)=\dfrac{4.29}{\epsilon \mu F}\sqrt{\dfrac{B+\epsilon F}{S \ T}}
    }
  \overfullrule 5pt
  \mathindent\linewidth\relax
  \advance\mathindent-259pt
  \end{equation}
where $F$  is the source flux, $\epsilon$  is the detector efficiency, $B$ is the background, $S$  is the collecting area and $T$  is the integration time and $\mu$ is the modulation factor. This last term is the amplitude of modulation for a 100$\%$ polarized signal in absence of background.

Thermal emission is not expected to be significantly polarized and its overlap on highly polarized non-thermal radiation, in the 15-20 keV energy band, should lead to low polarized signal. To detect some degree of polarization in this energy band, a polarimeter able to reach low MDP values is needed. On the contrary at higher energies, where thermal emission falls down, we expect high degree of polarization. Thus, a higher value of MDP may be enough to detect the polarimetric signals.

Since the GPD is a position sensitive detector, it can be placed behind a coded mask aperture.



\begin{figure}
\begin{center}
\includegraphics*[width = 6 cm, angle = 0]{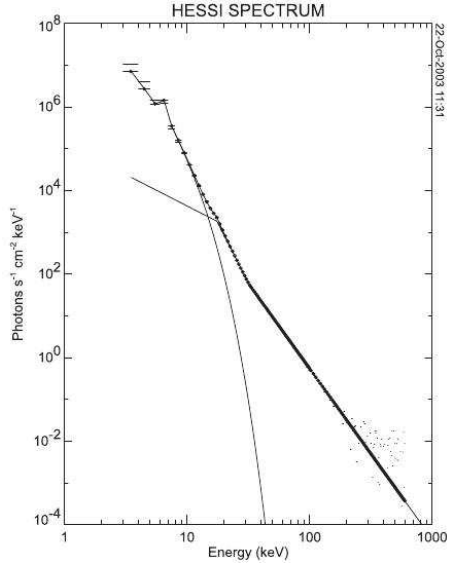}
\end{center}
\caption{The RHESSI X-ray spectrum of the hard X-ray source observed during the August 30, 2002 flare of class X 1.5 at 13:27:56 - 13:28:12 UT ($\Delta$t=16 s). The spectrum is the sum of the thermal and non-thermal components. \citep{Karl04}.
}\label{fig:figure6}
\end{figure}

In Table \ref{tab:table1} we report the MDP for the spectrum of Fig. \ref{fig:figure6} (integration time of 16 s) in the energy ranges 15-20 keV and 20-35 keV. 
The association with a coded mask leaves about a 50$\%$ of open area, corresponding about at 1 cm$^2$. The GPD is assumed to be filled with a gas mixture of argon (60$\%$) and DME (40$\%$) at the pressure of 3 bar for 3 centimetre of thickness of the gas cell. 

Polarized emission from thermal source may reach a degree of polarization up to about 3.3$\%$ at 25 keV for flare loop at the solar limb aligned with the line of site. 
Its contribution to total flux decreases rapidly at higher energies while the degree of polarization grows slightly up to about 4.3$\%$ at 50 keV \citep{EmsBro80}. At low energy the main contribution to total flux comes from such a low polarized thermal emission, for detecting which low value of MDP are needed.

Polarization of total emission in the HXRs is dominated by non-thermal emission. Close to the solar limb polarization degree of such a component may reach values of 30$-$40$\%$ at 20 keV, decreasing with energy down to about 20$\%$ at 200 keV. In the HXRs thermal emission can be negligible so that the main contribution to total flux is given by such a highly polarized non-thermal emission, for detecting which there is not a stringent requirement on the MDP.

From such considerations we have that the X-ray spectrum should be characterized by a variation on polarization properties moving from low to high energy. A low energy weakly polarized thermal emition should be followed by an highly polarized non-thermal emition.

Multiplying the detector area by a factor N, the MDP reduces of a factor $\sqrt{N}$ (see Eq.\eqref{eq:3}), thus looking at Table \ref{tab:table1} it is possible to conclude that the GPD may be employed for measuring X-ray polarization from solar flares for example if employed in an array configuration.

\begin{table}
\caption{Evaluation of the MDP  at the 99$\%$ confidence level in the energy ranges 15-20 keV and 20-35 keV for the spectrum reported in Fig.\ref{fig:figure6} ($\Delta$t=16 s)  and  for 1 cm$^2$  of open area. The GPD is assumed to be filled with a gas mixture of argon (60$\%$) and DME (40$\%$) at the pressure of 3 bar for 3 centimetre of thickness of the gas cell.}
\begin{tabular}{lll}
\hline
Energy Band  (keV)& Rate (counts/s) & MDP\\
\hline
15-20 & 1840 &5 $\%$ \\
20-35 & 230  &12 $\%$ \\
\hline
\end{tabular}
\label{tab:table1}
\end{table}

The angular resolution for a coded mask is given by
  \begin{equation}
    \label{eq:4}
 {%
   \Delta \alpha \approx  \arctan \left( \dfrac{l}{h}\right) 
    }
  \overfullrule 5pt
  \mathindent\linewidth\relax
  \advance\mathindent-259pt
  \end{equation}
  
where $l$ is the dimension of the mask elements and $h$ is the detector-mask separation. As an example, for mask elements of $l$=250 $\mu$m and a separation $h$= 2.5 m, the angular resolution is $\approx$20".

To increase significantly the collecting area for a detector, useful to reach higher MDPs in shorter integration times, it is possible to place the GPD at the focal plane of an X-ray telescope, taking advantage from the large effective area made available by the new multilayer technology, already developed for X-ray focusing up to 80 keV \citep{Par09, Tagl09}. This optic system, designed for faint celestial sources, is based on a classical Walter I geometry, made of 70 coaxial reflecting shells, with a focal length of 10 m and an angular resolution of about 20 '' at 30 keV.
One optic module has an effective area of about 200 cm$^2$ at 20 keV and reduces down to about 100 cm$^2$ at 35 keV \citep{Sof10}. 
Because of the GPD can work up to 35 keV, it is possible to reduce the number of the inner shells reflecting higher energies.
Moreover considering that this is a new technology, it may be possible to focus the effort to reach an  angular resolution better than 20''.


\section{Study of systematic effects}
The study of systematic effects is a key point for polarimeters developments, especially in the X-ray where few results have been acquired up to date, and they are very often controversial.

First of all the spurious modulation seen by the instrument detecting unpolarized radiation must be known. Having such a modulation as lower as possible is a good goal. Sometimes a spurious modulation depends on the peculiar geometry of the instrument \citep{Kraw2011}. Such an effect must be controlled for being confident of results.
The GPD demonstrated to reach very low spurious modulation levels. For gas mixture 20$\%$ He and 80$\%$ DME at 1 bar of pressure within 1 cm thick cell, it gives only a residual modulation of (0.18$\pm$0.14)$\%$ for on axis radiation of $^{55}$Fe at 5.9 keV \citep{Bel2009}.

Another effect to take into account is the blurring produced by inclined penetration and absorption of radiation within the thick gas cell, because photons are absorbed at different depths. If the instrument is coupled with a X-ray telescope, photons are absorbed before and after the optical focus, producing a blurring in the image, independently by the telescope PSF. We evaluated that such an effect will not compromise GPD imaging performance with respect to blurring produced by the telescope PSF \citep{Laz2009}. This effect is smaller for radiation beam approaching to the on axis direction.
Such a disturbing effect must be studied specifically for a coded mask coupling.

Grazing incident reflection of radiation on X-ray optics produces negligible effects on polarization \citep{Chi92, Chi93, Sanc93}, this is true also for multilayer optics systems \citep{Kat09}.

With respect to far astrophysical sources we expect to be more easily source dominated and the fast readout of the chip is an essential requirement for the electronics. This issue has been studied for the XPOL polarimeter on board of IXO satellite reaching 10 $\mu$s of dead time \citep{Bel09b} (b).

The wide characterization on ground which the GPD has undergone, allows us to consider such an instrument interesting for possible solar X-ray observations. However, because its original field of employment refers to galactic and extragalactic astrophysical sources, a deeper study of its application to study solar X-ray sources is needed and will be performed.

\section{Conclusion}

Polarimetry of solar flares is a diagnostics for the properties of the hot plasma in solar atmosphere. It's a tool to investigate the behaviour of the local magnetic field within regions where acceleration mechanisms of particles to high energies act. 

The GPD employed as solar X-ray polarimeter could help to investigate polarimetric properties of solar X-ray flares. 
It is able to provide source images while doing spectroscopy, timing and polarimetry.
It could be used in association with a coded mask aperture, necessary to obtain the image and localize the flares on the solar disc. A detector-mask separation of at least 2.5 m is needed to allow an angular resolution of at least 20 ''. The collecting area  is about 1 cm$^2$ for one detector. With this set up an array configuration is needed to achieve low MDP values. Significant results would be reached only for the most intense flares.

Otherwise the GPD can be employed also at the focal plane of an hard X-ray telescope built 
with the new multilayer technology. 
The angular resolution achievable at the present is about 20 '' at 30 keV.

Thus, a coded mask association may be employed to observe the most intense flares, otherwise the optic system is needed to observe less intense flares and in general the higher energy part of flares spectrum where the flux is very low.

If the capability of flares localization on the solar disc, and hence the angular resolution, are not necessary requirements, it is possible to use the GPD without coded aperture nor X-ray optics. In this case the flare localization should be achieved by other instruments, observing the flare at the same time. 




\end{document}